\documentclass[twocolumn,showpacs,preprintnumbers,amsmath,amssymb]{revtex4}
\usepackage{graphicx}
\usepackage{amsmath}
\usepackage{amsfonts}
\usepackage{amssymb}

\begin{document}

\title{Surface instabilities on liquid oxygen in an inhomogeneous magnetic field}
\author{A..T. Catherall, Keith A. Benedict, P. J. King, L. Eaves}
\affiliation{School of Physics and Astronomy, The University of Nottingham,
\\University Park, NOTTINGHAM, NG7 2RD, UK}
\date{\today}

\begin{abstract}
Liquid oxygen exhibits surface instabilities when subjected to a
sufficiently strong magnetic field. A vertically oriented magnetic field
gradient both increases the magnetic field value at which the pattern forms
and shrinks the length scale of the surface patterning. We show that these
effects of the field gradient may be described in terms of an ``effective
gravity'', which in our experiments may be varied from 1g to 360g.
\end{abstract}

\maketitle

During the course of levitation and magnetic flotation experiments, the
surface of liquid oxygen has been observed to exhibit an instability towards
corrugation when subjected to a sufficiently strong perpendicular magnetic
field \cite{Catherall}. Liquid oxygen (LOX) is strongly paramagnetic, a fact
that is simply explained by Hund's rule which dictates that the two
electrons in the anti-bonding $2\pi_{g}$ orbitals of the $O_{2}$ molecule
form a spin triplet. This strong paramagnetism leads to a number of
interesting properties of LOX when subjected to magnetic fields, such as:
magnetic flotation \cite{Catherall}; the magneto-volume effect\cite{Uyeda1},
in which the volume of the liquid changes under an applied field;
field-induced transparency \cite{Uyeda2}, in which the liquid loses its blue
colour.

For levitation and magnetic flotation, a large inhomogeneous magnetic field
is required \cite{Geim} and we observe that the presence of a vertical
magnetic field gradient strongly influences both the magnitude of the fields
required for the onset of this surface instability and also the
characteristic length scale of the patterns observed. It is the influence of
the field gradient which is the subject of this report.

Surface instabilities were first investigated in ferrofluids \cite
{Rosensweig,Rosensweig2} which consist of colloidal suspensions of
ferromagnetic particles (such as magnetite or cobalt) in organic liquids
(such as oil or kerosene) to which are added surfactants which coat the
particles in order to inhibit cohesion and prevent field-induced
flocculation.

\begin{figure}[h]
\begin{center}
\includegraphics[height=216pt, width=216pt, keepaspectratio]{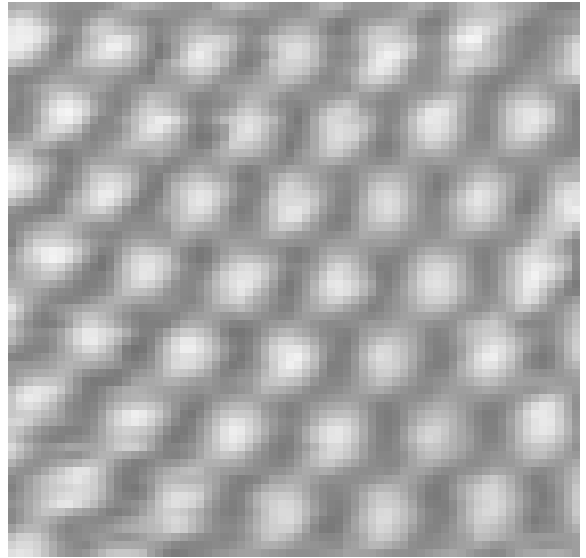}
\end{center}
\caption{View of the hexagonal patterns formed on the surface of liquid
oxygen held inside a glass Dewar within the bore of the magnet. The field of
view is $\sim 17mm\times17mm$.}
\label{triangular}
\end{figure}

\begin{figure}[h]
\begin{center}
\includegraphics [height=216pt, width=216pt, keepaspectratio]{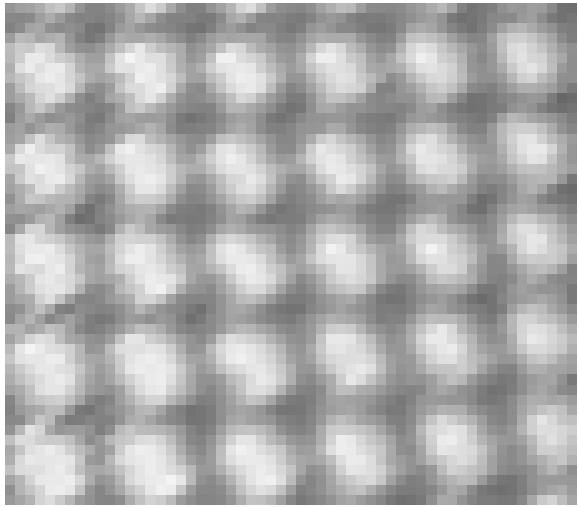}
\end{center}
\caption{View of square patterns formed on surface of LOX. The field of view
is $\sim 6.5mm \times 5.5mm$}
\label{square}
\end{figure}

If a sufficiently strong magnetic field is applied perpendicular to the
surface of a layer of magnetic fluid, that surface will develop static
corrugations \cite{Rosensweig,Rosensweig2}. The appearance of these surface
corrugations lowers the magnetic energy of the liquid because of the
focussing of the magnetic flux toward the peaks in the surface pattern. Such
corrugations, however, cost gravitational energy (in moving fluid from the
troughs to the peaks) and surface free energy (by increasing the total area
of the surface). Only when the gain in magnetic energy exceeds the cost in
gravitational and surface free energies will the surface spontaneously
corrugate. Above a lower critical field, the surface corrugations form a
triangular lattice. Figure \ref{triangular} 
 shows our own observation of such
a pattern on LOX. At higher fields, however, a square lattice is observed.
For our own observations of this on LOX see figure \ref{square}.

Rosensweig \cite{Rosensweig} carried out a linear stability analysis to
determine the critical field for the onset of the instability and the
wavelength of the corrugations at the onset, for the case of a ferrofluid
subject to a uniform field. This work was extended by Gailitis \cite
{Gailitis,Gailitis2} (see \cite{Friedrichs} and references therein for more
recent work) to include the first nonlinear corrections in order to examine
the criteria for pattern selection. The theory of surface instabilities in
the presence on inhomogeneous fields was developed by Zelazo and Melcher
\cite{Zelazo} and the principal effect of the field gradient can be regarded
as a renormalization of the acceleration due to gravity, $g\rightarrow
\widetilde{g}=g+f_{p}$ where $f_{p}$ is the force driving unit mass of
paramagnetic fluid towards the higher field region.

In this report we test the supposition that this is the principal effect of
the field gradient. Following a brief description of our experimental
arrangement we give a brief outline of the results of Zelazo and Melcher
\cite{Zelazo} as they apply to our system, along with specific predictions
for the critical applied field for surface corrugation and the wavelength of
the surface pattern at the onset of the instability. Finally we compare
these predictions with our experimental observations testing the validity of
the theory.

\begin{figure}[h]
\begin{center}
\includegraphics [width=226pt, keepaspectratio]{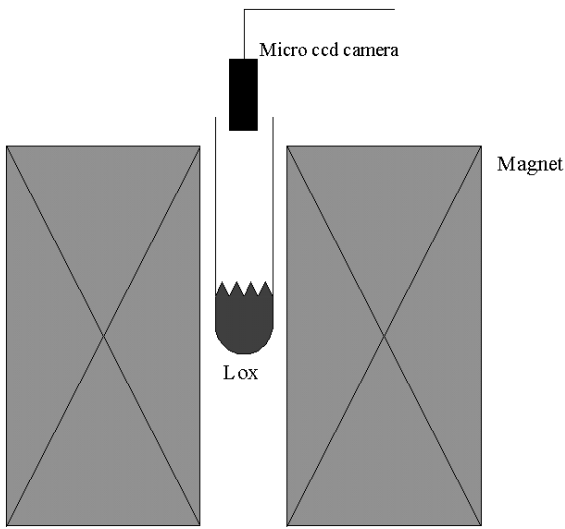}
\end{center}
\caption{Experimental set up}
\label{apparatus}
\end{figure}

The experiments were conducted using a specially constructed $17T$
superconducting Bitter-solenoid magnet having a $50mm$ diameter vertical
bore. For this magnet, the maximum value of $B_{0}dB_{0}/dz$ is $%
1470T^{2}m^{-1}$ offering an ``effective gravity'' of up to $360g$ for LOX.
A pool of LOX $1.5cm$ deep was contained in a glass Dewar vessel (see figure
\ref{apparatus}). 
The liquid was held at its boiling point at atmospheric pressure. It was
slowly lowered into the bore of the magnet, until a pattern of peaks was
observed. The magnet is designed to enhance radial variations in the
vertical field, so that the corrugation pattern first appears with perfect
triangular symmetry in the centre of the fluid surface. At higher fields the
pattern covers the entire surface but the circular container walls force
dislocations to appear in the regular lattice. The magnetic field and the
magnetic field gradient could be adjusted by the use of various magnet
currents and liquid surface positions.

A CCD camera was used to display a magnified image of the surface of the LOX
on a television screen, aiding the observation of the onset of surface
instabilities and facilitating more accurate measurements of the spatial
separation of the peaks.

The instability to surface corrugation only depends on the local properties of the field profile at
the surface of the liquid oxygen which can be characterized by the field at
the surface, $B_{0}$ and the length scale for field variations, $\zeta $,
defined by (recall that the field decreases with $z$ the distance from the centre of the magnet)
\begin{equation}
\zeta ^{-1}=\left( \frac{-1}{B\left( z\right) }\frac{\partial B\left(
z\right) }{\partial z}\right) _{\text{surface}}\qquad .
\end{equation}
The relevant material parameters are the density, $\rho $, the surface
tension, $\sigma $ and the susceptibility, $\chi $ of the oxygen. We take
these to have the following values at $90K$, $\rho =1149kgm^{-3}$, $\sigma =1.32\times
10^{-2}Nm^{-1}$ and $\chi =3.47\times 10^{-3}$\cite{CRC}.

Zelazo and Melcher \cite{Zelazo} carried out an analysis of the dynamical
behaviour of a ferrofluid surface subject to an inhomogeneous applied field.
The coupled equations for the magnetic field and the fluid were solved to
lowest order in the distortion of the surface to arrive at the dispersion
relation for surface modes. The situation in our experiments is simpler in
several respects than the general case considered in \cite{Zelazo}. Firstly,
the depth of the LOX layer and the height of the air space above it are both
much larger than the typical length scales for the observed corrugations so
that the fluid and the air above it may be considered to occupy infinite
half-spaces. Secondly, the magnetic response of the oxygen is essentially
linear over the whole range of fields that we consider, so that the
difference between $B/H$ and $\partial B/\partial H$ can be neglected. With
these simplifications the prediction of \cite{Zelazo} for the frequency of
the surface mode with wave number, $k$, is given by
\begin{equation}
\rho \omega _{k}^{2}=\sigma k^{3}-\lambda \frac{B_{0}^{2}}{\mu _{0}}%
k^{2}+\rho G\left( B_{0},\zeta \right) k \label{dispersion}
\end{equation}
where the coefficient
\begin{equation}
\lambda =\frac{\chi ^{2}}{2\left( 2+\chi \right) \left( 1+\chi \right) }\sim
\frac{\chi ^{2}}{4}\quad \chi \ll 1
\end{equation}
accounts for the energy gain (and the hence decrease in the restoring force) due to the
focusing of field lines through peaks in the surface and
\begin{equation}
G\left( B_{0},\zeta \right) =g\left( 1+\frac{\chi }{\rho g\left( 1+\chi
\right) \zeta }\frac{B_{0}^{2}}{\mu _{0}}\right)\label{renormalizedg}
\end{equation}
is the effective acceleration due to gravity, including the attractive force
on the diamagnetic oxygen towards the centre of the magnet.

Hence, a static distortion will appear with wavenumber
\begin{equation}
k=\frac{\lambda B_{0}^{2}\pm \sqrt{\lambda ^{2}B_{0}^{4}-4\mu _{0}^{2}\rho
\sigma G\left( B_{0},\zeta \right) }}{2\mu _{0}\sigma }  \label{wavenumber}
\end{equation}
provided the argument of the square root is positive. In consequence there is a
critical value of the field below which no such static distortion exists
\begin{equation}
B_{c}\left( \zeta \right) \approx \frac{8}{\chi }\sqrt{\frac{\mu _{0}\sigma
}{\chi \zeta }}\quad \chi \ll 1\quad .
\end{equation}
For comparison with the experiment it is useful to define $\widetilde{g}%
\left( \zeta \right) =G\left( B_{c}\left( \zeta \right) ,\zeta \right) $,
the value of the effective gravity at the onset field for a given field
gradient. The following relation should then hold
\begin{equation}
B_c^4=\frac{4\rho\sigma g}{\lambda^2} \left(\frac{\widetilde{g}_c}{g}\right) \label{Bcritical}
\end{equation}
The wavenumber of the static mode at the critical field is then
\begin{equation}
k_{c}=\frac{\lambda B_{c}^{2}}{2\mu _{0}\sigma }=\sqrt{\frac{\rho g}{\sigma }%
}\left( \frac{\widetilde{g}\left( \zeta \right) }{g}\right) ^{1/2}\approx
\frac{8}{\chi \zeta }\quad .
\end{equation}
The spacing between the peaks, in the triangular lattice observed, should
then be
\begin{equation}
L=\frac{4\pi }{\sqrt{3}k_{c}}=4\pi \sqrt{\frac{\sigma }{3\rho g}}\left(
\frac{\widetilde{g}\left( \zeta \right) }{g}\right) ^{-1/2}\qquad .
\label{criticalL}
\end{equation}

\begin{figure}[h]
\begin{center}
\includegraphics[width=240pt, keepaspectratio]{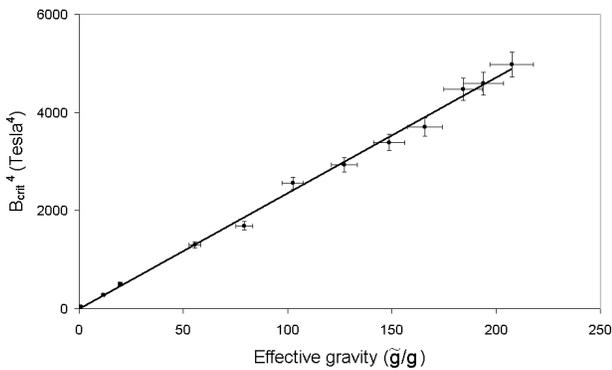}
\end{center}
\caption{The fourth power of the onset field for the hexagonal pattern, $%
B_{c}$, plotted against the effective gravity $\widetilde{g}/g$. The bars
indicate the accuracy of the data. The continuous line is a best fit to the
data.}
\label{data1}
\end{figure}

In order to test equation \ref{Bcritical} figure \ref{data1}
shows the experimental data presented as $B_{c}^{4}$ plotted
against effective gravity, $\widetilde{g}/g$. An estimate of the
experimental accuracy of each datum point is shown. The data are well fitted
by a straight line passing close to $\widetilde{g}/g=1$ at $B_{c}=0$ and
having a slope of $\left( 23.7\pm 0.8\right) \left( \text{Tesla}^{4}\right) $
confirming the expected dependence of $B_{c}$ upon $\widetilde{g}$. The
dominant error arises from the determination of the point of instability.
Using the values quoted above, equation \ref{Bcritical} predicts
\begin{equation}
B_{c}^{4}\approx \frac{\widetilde{g}}{g}\times 26.4\left( \text{Tesla}%
^{4}\right) \qquad .
\end{equation}
The discrepancy between theory and experiment is marginally significant and
will be the subject of further investigations.

\begin{figure}[h]
\begin{center}
\includegraphics[width=240pt, keepaspectratio]{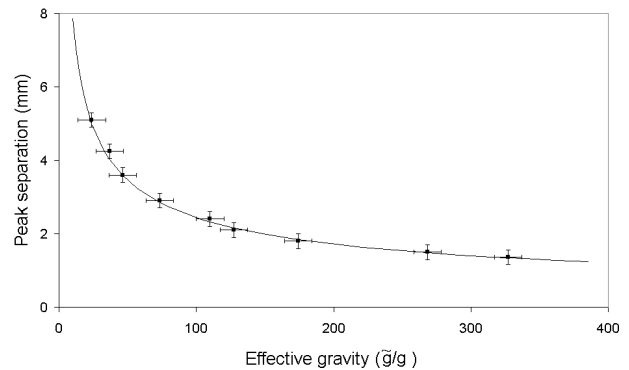}
\end{center}
\caption{The peak spacing $L$ for fields just above onset, plotted against
effective gravity $\widetilde{g}/g$. The continuous line is the prediction
derived from equation \ref{criticalL}, using the values for $\protect\rho$
and $\protect\sigma$ given in the text.}
\label{data2}
\end{figure}

The measured peak separation at onset, $L$, is plotted against $\widetilde{g}%
/g$ in figure \ref{data2},
together with the prediction of equation \ref
{criticalL} that $L^2\widetilde{g}/g=6.05\times10^{-4}m^2$ based on the
values of $\rho$ and $\sigma$ given above. The prediction fits the data very
well, a best fit to the data corresponding to $L^2\widetilde{g}%
/g=(6.2\pm0.3)\times10^{-4}m^2$.

We have observed the corrugation instability \cite{Rosensweig} in the
surface of liquid oxygen and examined the dependence of the critical field
and wavelength of the pattern on the field gradient. We have found that, in
agreement with the theory developed for ferrofluids, the principal effect of
a field gradient is to renormalize the gravitational acceleration in the
manner described by equation \ref{renormalizedg}, the experimental
observations offering satisfactory agreement with the theoretical
predictions for the onset field and the wavelength over the wide range of
effective gravities which we have examined. Liquid oxygen is not only a
homogeneous elemental magnetic liquid suitable for the study of fluid and
lattice dynamics over a very wide range of effective gravities, but also
offers a low cost environmentally friendly fluid for a range of
technological applications such as the separation of precious minerals.

\begin{acknowledgements}
The closed-cycle superconducting Bitter-Solenoid used in these experiments was
developed jointly by the School of Physics and Astronomy at the University of
Nottingham and Oxford Instruments with funding from EPSRC under the JREI
scheme. ATC is jointly funded by the EPSRC(UK) and Oxford Instruments.
\end{acknowledgements}

\end{document}